\documentclass[twocolumn]{IEEEtran}
\usepackage{amsmath}
\usepackage{latexsym}
\usepackage{amssymb}
\usepackage{bm}
\newtheorem{thm}    {Theorem}
\newtheorem{rem}     {Remark}

\def\complex{\mathbb{C}}
\newcommand{\defeq}{\stackrel{\rm def}{=}}

\newcommand{\cH}{{\cal H}}
\newcommand{\cS}{{\cal S}}
\def\choose#1#2{\genfrac{(}{)}{0pt}{}{#1}{#2}}
\def\cY{{\cal Y}}

\def\cD{{\cal D}}

\newcommand{\cP}{{\cal P}}

\newcommand{\cX}{{\cal X}}

\newcommand{\Tr}{{\rm Tr}\,}

\newcommand{\Pe}{{\rm P_e}}

\makeatletter
\newcommand{\lleq}{\mathrel{\mathpalette\gl@align<}}
\newcommand{\ggeq}{\mathrel{\mathpalette\gl@align>}}
\newcommand{\gl@align}[2]{
\vbox{\baselineskip\z@skip\lineskip\z@
\ialign{$\m@th#1\hfil##\hfil$\crcr#2\crcr{}_{{}_{(=)}}\crcr}}}
\makeatother

\newcommand{\ch}{W}

\newcommand{\charg}[1]{W_{#1}}
\newcommand{\chnarg}[1]{W^{(n)}_{#1}}

\newcommand{\coden}{\Phi^{(n)}}

\newcommand{\enc}{\varphi}

\newcommand{\encarg}[1]{\varphi(#1)}

\newcommand{\xn}{x^n}

\begin{document}
\title{Channel capacities of classical and quantum list decoding}
\author{
Masahito Hayashi
\thanks{
M. Hayashi is with ERATO-SORST Quantum Computation and Information Project, 
JST,
5-28-3, Hongo, Bunkyo-ku, Tokyo, 113-0033, Japan.
(e-mail: masahito@qci.jst.go.jp)}}
\date{}
\maketitle
\begin{abstract}
We focus on classical and quantum list decoding.
The capacity of list decoding was obtained by Nishimura in the case 
when the number of list does not increase exponentially.
However, the capacity of the exponential-list case is open even in 
the classical case
while its converse part was obtained by Nishimura.
We derive the channel capacities in the classical and quantum case
with an exponentially increasing list.
The converse part of the quantum case is obtained 
by modifying Nagaoka's simple proof for strong converse theorem 
for channel capacity.
The direct part is derived by a quite simple argument.
\end{abstract}
\section{Introduction}
\PARstart{L}{ist} 
decoding was introduced independently
by Elias \cite{Elias} and Wozencraft\cite{Wo} as relaxation of 
the notion of the decoding process.
In the list decoding, the decoder can choose more than one element
as candidates of the message sent by the encoder.
If one of these elements coincides with the true message,
the decoding is regarded as successful.
In this formulation, 
Nishimura \cite{Nish} obtained 
the channel capacity by showing its strong converse part\footnote{the strong converse part is the argument that the average error goes to $1$ if the code has a transmission rate over the capacity.}.
That is, he showed that
the transmission rate is less then the conventional capacity plus 
the rate of number of list.
Then, the reliable transmission rate does not increase
even if list decode is allowed
if the number of list does not increase exponentially.
The achievability of this bound has been proved only 
when the number of list is not exponentially increasing.
In the non-exponential case, 
these results was generalized by Ahlswede \cite{Ah}.

In this paper, we point out that 
the upper bound of capacity by Nishimura can be attained
even if the number of list increases exponentially.
Further, we treat the channel 
capacity of list decoding in a quantum setting.
Historically, its quantum version was treated by 
Kawachi \& Yamakami \cite{K-Y} from the viewpoint of
complexity theory, first.
However, they did not treat this problem as
the quantum extension from a viewpoint of
Shannon's communication theory.
Hence, we focus on the capacity of the classical-quantum channel\footnote{classical-quantum channel is a channel with classical input signals and 
quantum output states.}.
In this setting, 
the input quantum state is choosed dependently of the input classical 
message,
and sent it through a noisy quantum channel.
The receiver recovers the classical message
via a good quantum measurement.

On the other hand, 
Nagaoka \cite{Nag} obtained 
a quite simple proof of the strong converse part
of the classical capacity for classical channel and 
classical-quantum channel.
His proof extensively simplified the strong converse part
not only of the quantum case but also of the classical case.

As the main result, we extend 
Nishimura's result to the quantum setting.
That is, 
we show that the reliable transmission rate is less than 
the conventional capacity plus the rate of number of list
in quantum setting.
The proof is essentially based on 
a quite simple proof of converse part of quantum channel coding theorem 
by Nagaoka\cite{Nag}.
Thanks to simplicity of Nagaoka's proof,
we can simply prove the strong converse part.
Hence,
if we apply our proof to the classical case,
we obtain a simpler proof than existing proof of the strong converse part 
of list decoding\cite{Nish}.
Therefore, 
the discussion of this paper is meaningful 
for the classical viewpoint as well as 
the quantum viewpoint.
Thus, this paper is organized so that
the reader 
can understand the proof of the classical case
without any knowledge of the quantum case.

\section{Main results}
In the classical case, the channel is given 
by the output distribution of the output system
$\cY$ 
depending on the input signal $x$.
In the following, we describe this distribution by $W_x$.
Then, the relative entropy $D(W_x\|W_{x'})$ 
is given as
\begin{align*}
D(W_x\|W_{x'})&\defeq 
\sum_y W_x(y) \log W_x(y)- \log W_{x'}(y)
\end{align*}
A quantum extension of channel 
is given by a density matrix $W_x$
on the output system depending on $x$.
In this case, the relative entropy $D(W_x\|W_{x'})$ 
is given as
\begin{align*}
D(W_x\|W_{x'})&\defeq 
\Tr W_x (\log W_x- \log W_{x'})
\end{align*}
That is, $W_x$ is a distribution in the classical case, 
and it is a density matrix in the quantum case.
In these cases, the channel capacity $C(W)$ 
is given as\cite{Shannon48,HoCh,SW,Ohy-Pet-Wat,Sch-Wes:optimal}.
\begin{align}
&C(W)= 
 \max_{p\in \cP (\cX)} I(p,\ch) 
= 
\max_{p\in \cP (\cX)} \min_{\sigma \in \cS(\cH)}  
J(p,\sigma,W ) \nonumber \\
= &
\min_{\sigma \in \cS(\cH)}  
\max_{p\in \cP (\cX)} 
J(p,\sigma,W )
=
\min_{\sigma \in \cS(\cH)}  
\max_{x \in \cX}
D(W_x\|\sigma ) \label{7-15} ,
\end{align}
where
\begin{align}
I(p,W)  & \defeq   \sum_{x\in\cX} p(x) D(\charg{x}\|\charg{p}) 
,\label{11-5-30-q} \\
\charg{p} & \defeq  \sum_{x\in\cX} p(x) \charg{x},\label{11-5-31-q}\\
J(p,\sigma,W )&\defeq  \sum_{x \in \cX} p(x) D(W_x\|\sigma).
\end{align}

In this paper, we consider the capacity of the 
$L$-list decoding.
This problem is formulated as follows.
First, we fix the number $N$ corresponding to the size of the encoder.  
Next, choose $\varphi$ is a map, 
$\varphi : \{1, \ldots, N\} \rightarrow  \cX$, 
corresponding to the encoder.
Finally, we choose 
$\choose{N}{L}$ disjoint subsets $\cD
=(D_{(i_1, \ldots, i_L)})$
of $\cY$ in the classical case,
where $(i_1, \ldots, i_L)$ is the set of 
$L$ different elements $i_1, \ldots, i_L$.

In the quantum case, we choose 
$\choose{N}{L}$-valued POVM 
$M=\{M_{(i_1, \ldots, i_L)}\}$.
In the following, we call
the triplet $(N, \varphi, \cD)$ a classical $L$ list code, and 
call the triplet $(N, \varphi, M)$ a quantum $L$ list code.
For a classical $L$-list code 
$\Phi_L= (N, \varphi, \cD) $, we define the 
size $|\Phi_L |$ and the average error probability
$\Pe [\Phi_L]$ as
\begin{align*}
|\Phi_L |  &\defeq N , \\
\Pe [\Phi_L]
&\defeq 
\frac{1}{N}\sum_{i=1}^N 
\left(1- \sum_{j_1,\ldots, j_{L-1} \neq i }
\charg{\encarg{i}} \cD_{i,j_1,\ldots, j_{L-1}}\right)
\end{align*}
For a quantum $L$-list code 
$\Phi_L= (N, \enc, M) $, we define the 
size $|\Phi_L |$ and the average error probability
$\Pe [\Phi_L]$ as
\begin{align*}
|\Phi_L |  &\defeq N , \\
\Pe [\Phi_L]
& \defeq 
\frac{1}{N}\sum_{i=1}^N 
\left(1- \sum_{j_1,\ldots, j_{L-1} \neq i }
\Tr \charg{\encarg{i}} M_{i,j_1,\ldots, j_{L-1}}\right).
\end{align*}

Now, we can define the channel capacities of
classical and quantum list decoding.
Consider $ n $ communications.
For simplicity, let us assume that each communication is independent and 
identical.  
That is, the channel is given by the map
$W^{(n)}:\xn \defeq (x_1, \ldots , x_n) \mapsto \chnarg{\xn} \defeq
\charg{x_1}\times\cdots\times\charg{x_n}$ from the alphabet ${\cal X}^n$,
in the classical case.
and by 
$W^{(n)}:\xn \defeq (x_1, \ldots , x_n) \mapsto \chnarg{\xn} \defeq
\charg{x_1}\otimes\cdots\otimes\charg{x_n}$ from the alphabet ${\cal X}^n$,
in the quantum case.
In this case, an encoder of 
size $N_n$ is given by the map $\varphi^{(n)}$ from $\{1, \ldots ,N_n\}$ to 
${\cal X}^n$, and it is written as 
$\varphi^{(n)}(i)=(\varphi^{(n)}_1(i), \ldots, \varphi^{(n)}_n(i))$.  
Then, the capacity of $\{L_n\}$-list decoding
is given as
\begin{align}
C(W,\{L_n\})  &\defeq 
\sup_{\{\coden\}}
\left\{\left. \varliminf   \frac{1}{n} \log \frac{|\coden_{L_n} | }{L_n}
\right|
\lim   \Pe[\coden_{L_n}] =0 \right\}
\label{C-general} \\
C^\dagger(W,\{L_n\})  &\defeq 
\sup_{\{\coden\}}
\left\{\left. \varliminf   \frac{1}{n} \log \frac{|\coden_{L_n} | }{L_n}
\right|
\varliminf   \Pe[\coden_{L_n}] < 1 \right\}
\label{C-general2} 
\end{align}
\begin{thm}\label{21-2}
The equations 
\begin{align}
C(W,\{L_n\}) =C^\dagger(W,\{L_n\}) =C(W)
\end{align}
hold for any sequence $\{L_n\}$.
\end{thm}

Nishimura \cite{Nish} defined 
the capacity as $\sup_{\{L_n\}}C(W,\{L_n\})$.
He proved that $\sup_{\{L_n\}}C(W,\{L_n\})=\sup_{\{L_n\}}
C^\dagger(W,\{L_n\})=C(W)$
by combing the two fact $C(W,\{1\})= C(W)$ and 
$C^\dagger(W,\{L_n\}) \le C(W)$, which is the main result of his paper.
Ahlswede \cite{Ah} discussed the capacity $C(W,\{L_n\})$
only when $L_n$ is not exponentially increasing.
However, we can easily check that
$C(W,\{L_n\}) \ge C(W)$ for any sequence $\{L_n\}$ as follows.
Based on a usual code $(M_n, \phi, \cD)$,
we can construct a $L_n$-list code $(M_n L_n,\phi',\cD')$
as 
$\phi_{jL_n+ i}':=\phi_{j+1}$ for $0 \le i \le L_n$
and 
$D_{(j L_n+1,j L_n+2, \ldots, (j +1)L_n)}'=
D_{j+1}$.
Then, the error probability of 
$L_n$-list code $(M_n L_n,\phi',\cD')$ is equal to 
that of the code $(M_n, \phi, \cD)$.
Hence, we obtain 
the direct part $C(W,\{L_n\}) \ge C(W)$.
The quantum case also can be checked in a similar way.
Hence, it is sufficient to show the opposite 
inequality
$C^\dagger(W,\{L_n\}) \le C(W)$.

\begin{rem}
When $L_n$ does not increase exponentially,
we can show that
$C^\dagger(W,\{L_n\}) \le C(W)$
as follows\cite{Ma,Og}.
Let $\delta_n$ be the probability of correct decoding
of $L_n$-list decode.
When we randomly choose one element among $L_n$,
we obtain a conventional code with 
the probability $\frac{\delta_n}{L_n}$ of correct decoding.
From the strong converse theorem of conventional coding,
the value $\frac{\delta_n}{L_n}$ goes to $0$ exponentially.
Hence, the probability $\delta_n$ of correct decoding
also goes to $0$ exponentially.
Then, we obtain $C^\dagger(W,\{L_n\}) \le C(W)$.
However, its proof of the exponential-list case is more difficult.
Therefore, the strong converse of the exponential-list case 
is the main part of this paper.
\end{rem}

\section{Proof of Strong Converse Part}\label{s4.6}
In this section, we prove the strong converse parts 
by showing 
$C^\dagger(W,\{L_n\}) \le \min_{\sigma \in \cS(\cH)}  
\max_{p\in \cP (\cX)} 
J(p,\sigma,W )$.
For this purpose, we focus on 
the relative R\'{e}nyi entropy
and its monotonicity\cite{Csi70,Pe-Q}.
Its classical version is defined as 
$\phi(s|W_x\|W_{x'})\defeq\sum_{y}  (W_x(y))^{1-s}(W_{x'}(y))^s$,
and its quantum version as
$\phi(s|W_x\|W_{x'})\defeq\Tr W_x^{1-s}W_{x'}^s$.
We also define a channel version of the quantum relative R\'{e}nyi entropy
as $\phi(s|W\|\sigma) 
\defeq \max_{x \in \cX} \phi(s| W_x \|\sigma) $.

For a sequence of codes $\Phi_{L_n}^{(n)}$,
we 
choose a distribution/ density $\sigma$ such that 
\begin{align}
r\defeq \varliminf   \frac{1}{n}
\log | \Phi^{(n)}_{L_n}| \,> \max_{x \in \cX} D(W_x\|\sigma) \label{7-13} ,
\end{align}

As is shown later, 
the inequality 
\begin{align}
(1 - \Pe [\Phi^{(n)}_{L_n}])^{1-s}N_n^{-s}L_n^s
\le  e^{n \phi(s|W\|\sigma)} \label{6-14-9}
\end{align}
holds for $s \le 0$.
Thus,
\begin{align*}
\frac{1}{n}\log (1 - \Pe [\Phi^{(n)}_{L_n}])
\le \frac{\phi(s|W\|\sigma)+ \frac{s}{n} \log N_n- \frac{s}{n} \log L_n
}{1-s}.
\end{align*}
Letting
\begin{align}
r \defeq \varliminf   \frac{1}{n} \log N_n 
=\varliminf   \frac{1}{n} \log \frac{N_n }{L_n}
\label{19-1-1} ,
\end{align}
we obtain
\begin{align}
\varliminf
\frac{-1}{n}\log (1 - \Pe [\Phi^{(n)}_{L_n}])
\ge\frac{-sr - \phi(s|W\|\sigma)}{1-s}\label{ex-5-26-5}.
\end{align}
Reversing the order of the $\lim_{s \to 0}$ and $\max_{x \in \cX}$, we obtain
\begin{align}
& \phi'(0|W\|\sigma) =
\lim_{s \to 0} \max_{x \in \cX} 
\frac{\log \Tr W_x^{1-s}\sigma^{s}}{-s}  \nonumber \\
=& \max_{x \in \cX} \lim_{s \to 0} 
\frac{\log \Tr W_x^{1-s}\sigma^{s}}{-s} 
= \max_{x \in \cX} D(W_x \|\sigma)\label{koukan} .
\end{align}
Since $r > \max_{x \in \cX} D(W_x\|\sigma)$, 
we can choose a parameter $s_0 <0$ such that 
$\frac{{\phi}(s_0|W\|\sigma)- {\phi}(0|W\|\sigma)}{s_0} \,< r$.
Hence, we can show that 
\begin{align}
\frac{-s_0r - {\phi}(s_0|W\|\sigma)}{1-s_0}
= \frac{-s_0}{1-s_0}
\left( r - \frac{\phi(s_0|W\|\sigma)}{-s_0}\right)\,> 0 \label{7-19} .
\end{align}
Therefore, $1 - \Pe [\Phi^{(n)}_{L_n}]
\to 0$, and we obtain 
$C(W,\{L_n\}) \le \min_{\sigma \in \cS(\cH)}  
\max_{p\in \cP (\cX)} J(p,\sigma,W )$.

One may worry about the validity of reversing the order of $\lim_{s \to 0}$ and 
$\max_{x \in \cX}$ in (\ref{koukan}).  The validity of this step can be 
confirmed by showing that the
convergence is uniform with respect to $x$.  Since the dimension of our space is finite, 
$\{W_x \}_{x \in {\cal X}}$ is included in a compact set.  
The convergence with $s \to 0$, {\it i.e.}, 
$\frac{\log \Tr W_x^{1+s}\sigma^{-s}}{s} \to D(W_x \|\sigma)$, 
is uniform in any compact set, which shows the uniformity of the 
convergence.  
Therefore, we obtain (\ref{koukan}). 
\endproof

\subsection{Proof of (\ref{6-14-9}): Classical Case}
For a classical $L_n$-list code $\Phi_{L_n}^{(n)}
=(N_n,\varphi^{(n)},\cD^{(n)})$, 
we define distributions $R_n$ and $S_n$ on 
$\cY^n \times \{1, \ldots, N_n\}$
and subset $T_n$ of this set as follows:
\begin{align*}
S_n (y^n, i)& \defeq \frac{1}{N_n}\sigma(y^n)\\
R_n (y^n, i)& \defeq \frac{1}{N_n}
W^{(n)}_ {\varphi^{(n)}(i)} (y^n)\\
T_n  &\defeq \cup_{i}Y^{(n)}_{i} \times \{i\}
\end{align*}
where 
$Y^{(n)}_i= \cup_{j_1,\ldots, j_{L_n-1} \neq i }
D^{(n)}_{i,j_1,\ldots, j_{L_n-1}}$.
We have
\begin{align*}
R_n (T_n) =
\sum_{i=1}^{N_n} \frac{1}{N_n} 
W^{(n)}_{\varphi^{(n)}(i)}(Y^{(n)}_i)
= 1 - \Pe [\Phi^{(n)}_{L_n}] .
\end{align*}
On the other hand, 
for any element $y^n$, there is 
just $L_n$ inputs $i_1, \ldots, i_{L_n}$ such that
$y^n \in Y^{(n)}_{i_j}$.
Hence, we have
%
\begin{align}
&S_n (T_n) = \sum_{i=1}^{N_n} \frac{L_n}{N_n} 
\sigma^n (Y^{(n)}_i) =
 \frac{L_n}{N_n} \sigma^{n} (\cup_{i=1}^{N_n} Y^{(n)}_i )\nonumber\\
=& \frac{L_n}{N_n} \sigma^{n}(\cY^n ) 
=\frac{L_n}{N_n}.\label{3-4-2}
\end{align}
Note that this part is the main point of 
this paper. 
In other words, other parts are essentially parallel to 
Nagaoka's proof.
%
Using the monotonicity of 
relative R\'{e}nyi entropy\cite{Csi70},
we have
\begin{align*}
& R_n (T_n)^{1-s}
 S_n (T_n )^{s} \\
\le &
R_n (T_n )^{1-s}
S_n (T_n )^{s} 
+
R_n (T_n^c)^{1-s}
S_n (T_n^c)^{s}  \\
\le &
\sum_{(y^n,i)} R_n(y^n,i) ^{1-s} S_n (y^n,i)^{s} 
\end{align*}
for $s \le 0$.
Then,
\begin{align*}
& (1 - \Pe [\Phi^{(n)}_{L_n}])^{1-s}
N_n^{-s}L_n^s
=
R_n (T_n)^{1-s} S_n (T_n )^{s} \\
\le &
\sum_{(y^n,i)} R_n(y^n,i) ^{1-s} S_n (y^n,i)^{s} 
\nonumber \\
= &
\frac{1}{N_n} \sum_{i=1}^{N_n} 
\sum_{y^n} \left[(W^{(n)}_{\varphi^{(n)}(i)}(y^n))^{1-s}
(\sigma^{n}(y^n))^{s} \right]\\
= &
\frac{1}{N_n} \sum_{i=1}^{N_n} 
\prod_{l=1}^n 
\sum_y 
\left[(W_{\varphi^{(n)}_l(i)}(y))^{1-s}(\sigma(y))^{s} \right] 
\nonumber\\
\le & e^{n \phi(s|W\|\sigma)} .\nonumber
\end{align*}

\subsection{Proof of (\ref{6-14-9}): Quantum Case}
For a quantum $L_n$-list code $\Phi_{L_n}^{(n)}=
(N_n,\varphi^{(n)},M^{(n)})$, 
we define define density matrices $R_n$ and $S_n$ on $\cH^{\otimes n} \otimes 
\complex^{N_n}$ and a matrix $T_n$ as follows:
\begin{align*}
S_n & \defeq \frac{1}{N_n}
\left( \begin{array}{ccc}
\sigma^{\otimes n} & &  \smash{\lower1.4ex\hbox{0}}
\\
\smash{\lower1.7ex\hbox{0}} & \ddots & \\
&& \sigma^{\otimes n}
\end{array}
\right)  , \\
R_n & \defeq \frac{1}{N_n}
\left( \begin{array}{ccc}
W^{(n)}_ {\varphi^{(n)}(1)} & &  \smash{\lower1.4ex\hbox{0}}
\\
\smash{\lower1.7ex\hbox{0}} & \ddots & \\
&& W^{(n)}_ {\varphi^{(n)}(N_n)} 
\end{array}
\right) ,\\
T_n  & \defeq
\left( \begin{array}{ccc}
Y^{(n)}_1 & &  \smash{\lower1.4ex\hbox{0}}
\\
\smash{\lower1.7ex\hbox{0}} & \ddots & \\
&& Y^{(n)}_{N_n} 
\end{array}\right)  ,
\end{align*}
where 
$Y^{(n)}_i= 
\sum_{j_1,\ldots, j_{L_n-1} \neq i }
M_{i,j_1,\ldots, j_{L_n-1}}$.
Since $I \ge T_n \ge 0$, we have
\begin{align*}
\Tr R_n T_n =
\sum_{i=1}^{N_n} \frac{1}{N_n} \Tr W^{(n)}_{\varphi^{(n)}(i)}Y^{(n)}_i
= 1 - \Pe [\Phi^{(n)}_{L_n}].
\end{align*}
On the other hand, in the summation $\sum_{i=1}^{N_n} Y^{(n)}_i $,
we add the matrix
$M_{i,j_1,\ldots, j_{L_n-1}}$, $L_n$ times.
Hence, we have
\begin{align}
L_n I = \sum_{i=1}^{N_n} Y^{(n)}_i ,\label{3-4-1}
\end{align}
which implies 
\begin{align*}
&\Tr S_n T_n = \sum_{i=1}^{N_n} \frac{L_n}{N_n} \Tr \sigma^{\otimes n}
Y^{(n)}_i \\
=& \frac{L_n}{N_n} \Tr \sigma^{\otimes n} \sum_{i=1}^{N_n} Y^{(n)}_i 
= \frac{L_n}{N_n} \Tr \sigma^{\otimes n} =\frac{L_n}{N_n}.
\end{align*}
Note that this part is the main point of 
this paper. 
In other words, other parts are essentially parallel to 
Nagaoka's proof.
%
Using the monotonicity of 
quantum relative R\'{e}nyi entropy\cite{Pe-Q},
we have
\begin{align*}
&(\Tr R_n T_n )^{1-s}
(\Tr S_n T_n )^{s}  \\
\le &
(\Tr R_n T_n )^{1-s}
(\Tr S_n T_n )^{s} \\
&+
(\Tr R_n (I-T_n) )^{1-s}
(\Tr S_n (I-T_n) )^{s} \\
\le &
\Tr R_n ^{1-s} S_n ^{s} 
\end{align*}
for $s \le 0$.
Then,
\begin{align}
& (1 - \Pe [\Phi^{(n)}_{L_n}])^{1-s}
N_n^{-s}L_n^s
=
(\Tr R_n T_n )^{1-s}
(\Tr S_n T_n )^{s}  \nonumber \\
\le &
\Tr R_n ^{1-s} S_n ^{s} 
= 
\frac{1}{N_n} \sum_{i=1}^{N_n} 
\Tr \left[(W^{(n)}_{\varphi^{(n)}(i)})^{1-s}(\sigma^{\otimes n})^{s} \right]
\nonumber \\
=&
\frac{1}{N_n} \sum_{i=1}^{N_n} 
\prod_{l=1}^n 
\Tr \left[(W_{\varphi^{(n)}_l(i)})^{1-s}\sigma^{s} \right] 
\le  e^{n \phi(s|W\|\sigma)}. \nonumber
\end{align}

\section{Concluding remark}
The main point of Nagaoka's proof is the reduction of strong converse part of
channel capacity to hypothesis testing problem.
Hence, the essential point of this paper is
linking the strong converse part of the capacity of the list decoding to 
the hypothesis testing.
This relation is essentially given in (\ref{3-4-2}) and (\ref{3-4-1}).
Further,
as is mentioned in Hayashi \& Nagaoka \cite{Hay-Nag} and Hayashi \cite{Book},
Nagaoka's simple proof can be extended to capacity theorem 
with cost constraint.
Combining (\ref{3-4-1}) and (\ref{3-4-2}), we can easily obtain the 
capacity for list decoding with cost constraint.

Moreover, 
the capacity of the general sequence of channels 
was also derived in the classical case \cite{V-H} and in 
the quantum case \cite{Hay-Nag}.
The converse part is essentially derived by 
linking this problem to the hypothesis testing \cite{Hay-Nag}.
Hence, using formulas (\ref{3-4-2}) and (\ref{3-4-1}),
we can expect the same formula for list decoding.

\section*{Acknowledgments}
The author would like to thank Professor Hiroshi Imai of the
QCI project for support.
He is grateful to Dr. Tomoyuki Yamakami for 
useful discussions.
He is also grateful to Professor Andreas Winter for giving 
important information concerning the manuscript \cite{Ah}.
He also benefited from discussions with Professor Keiji Matsumoto and 
Dr. Tomohiro Ogawa.

\bibliographystyle{IEEE}

\end{document}